\documentstyle[prc,multicol,aps,psfig]{revtex} 

\def\bea{\begin{eqnarray}}
\def\eea{\end{eqnarray}}
\def\be{\begin{equation}}
\def\ee{\end{equation}}

\def\p{\prime}

\begin{document}
\title{Observables and initial conditions
for self-similar ellipsoidal flows}
\author{T. Cs{\"o}rg{\H o}$^{1,2}$, S.V. Akkelin$^{3}$,
Y. Hama$^{2}$, B. Luk\'acs$^1$ and Yu.M.  Sinyukov$^{2,3}$}
\address{$^1$MTA KFKI RMKI, H-1525 Budapest 114, POB 49, Hungary\\
$^2$ IF-USP, C.P. 66318, 05389-970 S\~ao Paulo, SP, Brazil,\\
$^3$ Bogolyubov Institute for Theoretical Physics, Kiev 03143,
Metrologichna 14b,Ukraine}
\maketitle

\begin{abstract}
Single-particle spectra and two-particle Bose-Einstein correlation 
functions are determined analytically utilizing a self-similar
solution of non-relativistic hydrodynamics for
ellipsoidally-symmetric, expanding fireballs, by assuming that the
symmetry axes of the ellipsoids are tilted in the frame of the
observation. The directed, elliptic and third flows are calculated
analytically. The mass dependences of the slope parameters in the
principal directions of the expansion, together with the mass and
angular dependences of the HBT radius parameters, reflect directly
the ellipsoidal properties of the flow.
\end{abstract}

\begin{multicols}{2}
{\it Introduction} --- The equations of hydrodynamics reflect the 
local conservation of matter, momentum and energy. These equations 
are well suited to study problems related to flows in various 
fields ranging from evolution of galaxies in astrophysics to 
heavy-ion and elementary-particle collisions in high-energy 
physics. The finding of exact self-similar hydro solutions 
sometimes represents essential progress in physics, as the 
discovery of the Hubble flow of our Universe or the Bjorken flow 
of ultrarelativistic heavy-ion collisions.

In this Letter we consider the case of a non-relativistic 
hydrodynamical problem with ellipsoidal symmetry. Our goal is to 
demonstrate the influence of initial conditions on the final state 
observables, utilizing an explicit, exact  and simple analytic 
solution of fireball hydrodynamics. In particular, we attempt to 
understand the relationship between  the initial conditions (the 
ellipsoidal asymmetry  and the tilt of the major axis) and the 
final observables. 

{\it The new family of self-similar ellipsoidal solutions} --- A
self-similar solution of non-relativistic hydrodynamics with ideal 
gas equation of state and a generalized (direction-dependent) 
Hubble flow, a three-dimensional ellipsoidal Gaussian density 
profile and a homogeneous, space-indepenent temperature profile 
has just been found in ref.~\cite{ellsol}. This solution has many 
interesting properties, e.g. the partial differential equations of 
hydrodynamics are reduced to ordinary differential equations 
corresponding to a Hamiltonian motion of a massive particle in a 
non-central repulsive potential. These results correspond to the 
generalization of earlier, data motivated hydrodynamical 
parameterizations and/or solutions of 
refs.~\cite{jnr,nr,cspeter,cssol,nrt,3d,jde} to ellipsoidal 
symmetry and non-central heavy-ion collisions with homogeneous
temperature profile.

Below we generalize the hydrodynamic solution of ref.~\cite{ellsol} 
for some fairly wide family of thermodynamically consistent 
equations of state, and calculate analytically all the observables 
of non-central collisions. 
It allows, in principle, to solve an inverse problem, namely, 
given an (in general non-ideal) equation of state, to restore the 
initial conditions from the observables.
However, we do not aim here to apply directly the new hydro
solution to data fitting in high-energy heavy-ion physics. In
order to reach the level of data fitting, generalizations to
relativistic flow patterns, more realistic equations of state and
temperature profiles are needed. Some of these generalizations
seem to be straightforward and are in progress~\cite{nrtgen,1drel}. 

Consider the non-relativistic hydrodynamical problem, as given by
the continuity, Euler and energy equations:
\begin{eqnarray}
{\partial_t\,n} + \nabla\cdot(n{\bf v}) & = & 0\,,
 \label{e:cont} \\
{\partial_t\,{\bf v}} + ({\bf v}\cdot{\bf\nabla}){\bf v} & = & 
 - ({\bf \nabla} p) / (m n)\, ,  \label{e:Eu} \\
{\partial_t\,\epsilon} + {\bf\nabla}\cdot(\epsilon{\bf v}) & = & 
 - p {\bf \nabla}\cdot{\bf v}\, ,  \label{e:en}
\end{eqnarray}
where $n$ denotes the particle number density, {\bf v} stands for 
the non-relativistic (NR) flow velocity field, $\epsilon$ for the 
NR energy density, $p$ for the pressure and in the following the 
temperature field is denoted by $T$. This set of equations are 
closed by some equation of state (EoS). We have chosen 
analytically solvable generalization of NR ideal gas equations of 
state: 
\begin{eqnarray} 
p & = &  n T\,, \qquad \epsilon = \kappa (T) n T ,\label{e:eos} 
\end{eqnarray} 
which allow to study the solutions of NR hydro equations for any
temperature dependent ratio of pressure to energy density, 
$ p/\epsilon =1/ \kappa (T) $. The EoS (\ref{e:eos}) are
thermodynamically consistent for any function $\kappa (T)$, as 
can be checked by using the free energy density $f(T,n)$ and the 
relations 
\begin{eqnarray}
p & = & n \frac{\partial f}{\partial n} -f\,, \qquad \epsilon 
 = f - T \frac{\partial f}{\partial T}\ .\label{e:freeen}
\end{eqnarray}
The function $\kappa(T)$ characterizes the $p/\epsilon$ ratio 
for a broad variety of materials: e.g. a non-relativistic ideal 
gas yields $\kappa(T) = 3/2$. Note, that for finite-size systems 
phase transition can occupy certain temperature interval similar 
to a crossover. Then one can model such a change of the pressure
to energy density ratio at phase transition from deconfined 
quark matter to hadronic one by means of smooth variations of 
the values of $\kappa (T)$ in certain temperature domain.
Note also that it is usual to introduce the speed of sound
as $c_s^2 = dp/d\epsilon = 1/\kappa(T)$, so we model the 
change in the equation of state  essentially with the help of 
a temperature dependent speed of sound. 

For reasons of convenience we choose $n$, ${\rm{\bf v}}$ and $T$
as the independent hydrodynamic variables. 
The NR hydro equations are solved, similarly as it was done for 
the case of NR ideal gas EoS in ref. \cite{ellsol}, by the 
following self-similar, ellipsoidally symmetric density and flow 
profiles:
\begin{eqnarray}
n(t,{\bf r}^\p) & = & n_0 {\frac{V_0}{V}}\, 
 \mathrm{exp}\left({-{\frac{r_x^{\p\, 2}}{2X^2}}
                    -{\frac{r_y^{\p\, 2}}{2Y^2}}
                    -{\frac{r_z^{\p\, 2}}{2Z^2}}}
             \right), \label{n} \\
{\bf v}^\p(t,{\bf r}^\p) &=& \left({\frac{\dot{X}}{X}}\, r_x^\p,
                                   {\frac{\dot{Y}}{Y}}\, r_y^\p,
                                   {\frac{\dot{Z}}{Z}}\, r_z^\p 
                             \right), \label{v}
\end{eqnarray}
where the variables are defined in a center of mass frame $K^\p$,
but with the axes pointing to the principal directions of the
expansion. The time dependent scale parameters are denoted by
$(X,Y,Z)=(X(t),Y(t),Z(t))$, the typical volume of the expanding
system is $V = XYZ$, and the initial temperature and volume are
$T_0=T(t_0)$ and $V_0=V(t_0)$, and $n_0$ is a constant. The time
evolution of the radius parameters $X$, $Y$, $Z$ and temperature
$T$ are governed by the ordinary differential equations
\begin{eqnarray}
\ddot{X}X=\ddot{Y}Y=\ddot{Z}Z=\frac{ T}{m}\,, \label{scales}
\\ \dot{T}\frac{d}{dT}(\kappa T) + T\left({\frac{\dot{X}}{X}}+
     {\frac{\dot{Y}}{Y}}+
     {\frac{\dot{Z}}{Z}} \right)=0\,.
\label{ScT}
\end{eqnarray}
Note that the equation for the time dependence of the temperature
can be integrated in a straigthforward manner to find 
\begin{equation}
\frac{V_0}{V} = \exp\left[\kappa(T) - \kappa(T_0)\right]
	          \exp\int_{T_0}^T \frac{dT^\prime}{T^\prime}
		    \kappa(T^\prime)\,, 
\end{equation}
and this equation further simplified, in the case of a 
temperature independent $\kappa$, as
\begin{equation}
	T = T_0 \left(\frac{V_0}{V}\right)^{1/\kappa}.
\end{equation}

{\it Observables from the new solution} --- In order to evaluate
the measurable quantities, any hydrodynamical solution has to be
supplemented with an additional freeze-out criterion, that
specifies the end of the hydrodynamical evolution. Here we assume
sudden particle freeze-out at a constant temperature 
$T(t_f,{\bf r}) = T_f\,$ where EoS corresponds approximately to 
ideal gas ($\kappa (T_{f})=3/2$). 
This freeze-out condition is reached everywhere at the same time 
in the considered class of exact hydrodynamical solutions and it
is motivated by the simplicity of the results. Then, the emission
function is proportional to
\be
S(t,{\bf r}^\p,{\bf k}^\p) \propto
{\rm e}^{- \frac{ ({x\bf k}^\p - m {\bf v}^\p)^2}{2 m T_f}
         -{\frac{r_x^{\p\, 2}}{2X_f^2}}
         -{\frac{r_y^{\p\, 2}}{2Y_f^2}}
         -{\frac{r_z^{\p\, 2}}{2Z_f^2}}}\, \delta(t - t_f)\,. 
 \label{e:sxp}
\ee

{\it Single particle spectrum} ---
 The single-particle spectrum and the two-particle correlation
function can be evaluated similarly to that of
ref.~\cite{nr,cspeter}: 
\begin{eqnarray}
E\frac{d^3n}{d{\bf k}^\p} & \propto &
      E\exp\left(-\frac{k_x^{\p\, 2}}{2 m T_x^\p}
                 -\frac{k_y^{\p\, 2}}{2 m T_y^\p}
                 -\frac{k_z^{\p\, 2}}{2 m T_z^\p}
           \right),               \label{e:ellsp}\\
T_x^\p & = & T_f + m \dot X_f^2\ , \label{e:txp}\\
T_y^\p & = & T_f + m \dot Y_f^2\ , \label{e:typ}\\
T_z^\p & = & T_f + m \dot Z_f^2\ . \label{e:tzp}
\end{eqnarray}
Here $E=m+{\bf k}^{\p\,2}/(2m)$ in the non-relativistic limit we 
are considering, ${\bf k}^\p = (k_x^\p, k_y^\p, k_z^\p)$ stands 
for the momentum vector in $K^\p$, $X_f = X(t_f)$, etc. In the 
spherically symmetric case of $X = Y = Z = R$, we recover the 
earlier results~\cite{nr,cspeter}, with 
$\langle u \rangle = \dot R$ and 
$T_{\rm eff} = T_f + m \langle u \rangle^2$. 

The observables are determined in the center of mass frame of the
collision, $K$, where the $r_z$ axis points to the direction of 
the beam and the $r_x$ axis to that of the impact parameter.
In this frame, the coordinates and the momenta are denoted by
${\bf x}$ and ${\bf k}$.
We assume that the initial state of the hydrodynamic evolution
corresponds to a rotated ellipsoid in $K$. The tilt angle 
$\theta$ represents the rotation of the major (longitudinal) 
direction of expansion, $r_z^\p$ from the beam axis $r_z$. Hence 
the event plane is the $(r_x^\p, r_z^\p)$ plane, which is the 
same as the $(r_x,r_z)$ plane. The (zenithal) angle between 
directions $r_z$ and $r_z^\p$ is $\theta$, while the (azimuthal) 
angle between the transverse momentum ${\mathbf k}_t$ and the 
event plane is $\phi\,$.

The ellipsoidal spectrum of eq.~(\ref{e:ellsp}) generates the
following $\phi$ averaged single-particle spectrum in the $K$ 
frame:
\begin{eqnarray}
 \frac{d^2n}{2\pi k_t dk_t dk_z} & \propto &
   \exp\left(-\frac{k_t^2}{ 2 m T_{\rm eff}}
             -\frac{k_z^2}{ 2 m T_z}
       \right) f(v,w)\ ,  \label{e:ellsptr}\\
 \frac{1}{T_z} & = &
   \frac{\cos^2 \theta}{T_z^\p} +
   \frac{\sin^2 \theta}{T_x^\p}\ ,\\
 \frac{1}{T_x} & = &
   \frac{\cos^2 \theta}{T_x^\p} +
   \frac{\sin^2 \theta}{T_z^\p}\ ,\\
 \frac{1}{T_{\rm eff}} & = &
   \frac{1}{2}\left(\frac{1}{T_x} + \frac{1}{T_y^\p} \right),\\
 w & = & \frac{k_t^2}{4 m}
           \left( \frac{1}{T_y^\p} - \frac{1}{T_x} \right),\\
 v & = & - \frac{k_t k_z} {2m} \sin(2\theta)
           \left(\frac{1}{T_x^\p} - \frac{1}{T_z^\p} \right),\\
 f(v,w)& \approx & I_0(w) + \frac{v^2}{4}
   \left[I_0(w)+I_1(w)\right]\ ,
\end{eqnarray} 
where $f(v,w)$ is calculated for $|v| \ll 1$ and
$I_n(w) = \frac{1}{\pi} \int_0^\pi dz \cos(n z) \exp[w \cos(z)]$
is the modified Bessel function of order $n$ ($n=0,1,...)$. For
small ellipsoidal asymmetries, $w \ll 1$, $I_0(w) \simeq 1$ and
the effective temperature parameter in the transverse direction 
is the harmonic mean of the temperature parameters of the 
principal directions of expansion (projected to the transverse 
plane). As $T_z^\p \ge T_x^\p \ge T_y^\p$ is expected from the 
initial conditions, we obtain $T_x \ge T_x^\p$ and 
$T_z\le T_z^\p\,$.

The flow coefficients $v_n$ are defined as
\begin{equation}
 \frac{d^3n}{dk_z k_t dk_t d\phi} =
 \frac{d^2n}{2 \pi dk_z k_t dk_t}
      \left[1 + 2 \sum_{n=1}^{\infty} v_n \cos(n\phi)\right]. 
 \label{e:harm}
\end{equation}
Here $v_1$ is called the directed flow, $v_2$ the elliptic flow 
and $v_3$ the third flow. The transverse- and longitudinal- 
momentum dependence of the $v_n$ flow components can be written 
in terms of $v$ and $w$. Assuming that the tilt angle $\theta$ 
or the anisotropy is small, $|v| \ll 1$, the directed, elliptic 
and third flow components are evaluated as
\begin{eqnarray}
 v_1 & = & \frac{v}{2}\left[1+\frac{I_1(w)}{I_0(w)}\right],
           \label{e:v1} \\
 v_2 & = & \frac{I_1(w)}{I_0(w)}
       + \frac{v^2}{8}\left[ 1 + \frac{I_2(w)}{I_0(w)} -
                            2\left(\frac{I_1(w)}{I_0(w)}\right)^2
                      \right], \label{e:v2} \\
 v_3 & = & \frac{v}{2}\,\frac{I_2(w) + I_1(w)}{I_0(w)}\ . 
       \label{e:v3}
\end{eqnarray}
An angular tilt $\theta \ne 0$ is evidenced by the rise of the
directed and third flows as a function of rapidity 
$y = 0.5\ln [(E + k_z)/(E-k_z)]$ and by a minimum of the elliptic 
flow at mid-rapidity, see Fig.~1. This and other features are in
qualitative agreement with most of the data on intermediate- and 
high-energy heavy-ion 
collisions~\cite{dflow,na49flow,starflow,3flow}, suggesting that 
in non-central collisions the dominant longitudinal direction of 
expansion is slightly deviating from the beam direction. A more 
straightforward proof of the ellipsoidal nature of the flow can 
be obtained by determining the mass dependence of the parameters 
$T_x^\p\,$, $T_y^\p$ and $T_z^\p\,$, cf. 
eqs.~(\ref{e:txp}-\ref{e:tzp}) and Fig.~2.

{\it Two-particle correlations} --- The two-particle 
Bose-Einstein correlation function (BECF) is related to a 
Fourier-transform of the emission (or source) function 
$S(t,{\bf r}^\p,{\bf k}^\p)$ of eq.~(\ref{e:sxp}), see e.g. 
refs.~\cite{nr,cspeter,nrt}. If the core-halo 
picture~\cite{chalo} is valid, an effective intercept parameter 
$\lambda\equiv\lambda({\bf k}) = [N_c({\bf k})/N({\bf k})]^2$ 
appears, that measures the fraction of particles emitted directly 
from the core. The two-particle BECF is diagonal in $K^\p$, as
\begin{eqnarray}
 C({\bf K}^\p,{\bf q}^\p) & = & 1 +
    \lambda \exp\left(- q_x^{\p\, 2} R_x^{\p\, 2}
                      - q_y^{\p\, 2} R_y^{\p\, 2}
                      - q_z^{\p\, 2} R_z^{\p\, 2}
                \right), \label{e:ellbecf}\\
 {\bf K}^\p & = & {\bf K}^\p_{12}\, =
                  \, 0.5({\bf k}^\p_1 + {\bf k}^\p_2)\,, \\
 {\bf q}^\p & = & {\bf q}^\p_{12}\, = 
                  \, {\bf k}^\p_1 - {\bf k}^\p_2 \, =
                  \, (q_x^\p\,, q_y^\p\,, q_z^\p)\,, \\
 R_x^{\p\, -2} & = & X_f^{-2} \left( 1 + \frac{m}{T_f} \dot X_f^2
                              \right), \label{e:rxp}\\
 R_y^{\p\, -2} & = & Y_f^{-2} \left( 1 + \frac{m}{T_f} \dot Y_f^2
                              \right), \label{e:ryp}\\
 R_z^{\p\, -2} & = & Z_f^{-2} \left( 1 + \frac{m}{T_f} \dot Z_f^2
                              \right). 
\label{e:rzp}
\end{eqnarray}
These radius parameters measure the lengths of
homogeneity~\cite{sinyukov}. They are dominated by the shortest 
of the geometrical scales $(X_f, Y_f, Z_f)$ and the corresponding 
thermal scales defined by $(X_T, Y_T, Z_T )=\sqrt{\frac{T_f}{m}}
(\frac{X_f}{\dot X_f},\frac{Z_f}{\dot Z_f},\frac{Z_f}{\dot Z_f})$,
generalizing the results of refs.~\cite{nr,nrt,3d} to ellipsoidal
flows. The geometrical scales characterize the spatial variation 
of the fugacity term $\mu(t,{\bf r}^\p)/T(t,{\bf r}^\p)$, while 
the thermal scales characterize the spatial variations of the 
Boltzmann term $E_{{\rm loc}}^\p(t,{\bf r}^\p)/T(t,{\bf r}^\p)$, 
both of them evaluated at the point of maximal emittivity. In the 
$K^\p$ frame, cross-terms~\cite{xterm} vanish, $R^2_{i\ne j} = 0$, 
if the emission is sudden.

If the particle emission is gradual, but it happens in a narrow
interval $\Delta t$ centered at $t_f\,$, then the BECF can be
evaluated using the replacement 
$\delta(t-t_f)\rightarrow(2\pi\Delta t^2)^{-1/2}\exp[-(t-t_f)^2/
2\Delta t^2]$ in eq.~(\ref{e:sxp}), if $\Delta t\dot X_f<<X_f\,$, 
etc. Hence all the previous radius components, including the 
cross-terms, are extended with an additional term 
$\delta R^{\p\, 2}_{ij} = \beta_i^\p \beta_j^\p \Delta t^2$, 
where $\mbox{\boldmath$\beta$}^\p=({{\bf k}^\p_1+{\bf k}^\p_2})/
({E^\p_1 + E^\p_2})$ is the velocity of the pair in $K^\p$. 

The BECF's are usually given in the side-out-longitudinal or 
Bertsch-Pratt (BP) parameterization. The longitudinal direction, 
$r_{\rm long}\equiv r_{\rm l}$ in BP coincides with the beam 
direction. The plane orthogonal to the beam is decomposed to a 
direction parallel to the mean transverse momentum of the pair, 
$r_{{\rm out}}\equiv r_{\rm o}\,$, and the one perpendicular both 
to this and the beam direction, $r_{\rm side} = r_{\rm s}\,$. The 
mean velocity of the particle pair can be written in BP as 
$\mbox{\boldmath $\beta$}=(\beta_{\rm o},0,\beta_{\rm l})$, where 
$\beta_{\rm o} = \beta_t\,$. Let $\phi$ denote the angle of the 
event plane and the mean transverse momentum of the measured 
pair. The result is
\begin{eqnarray}
 C_{2}({\bf K},{\bf q}) &=&1+\lambda \exp
      \left(-\sum_{i,j={\rm s,o,l}}q_i q_j R_{ij}^2\right), \\
 R_{\rm s}^{2} &=& R_{y}^{\p\, 2} \cos^{2}\phi
                + R_{x}^{2} \sin^{2}\phi\,, \\
 R_{\rm o}^{2} &=& R_{x}^{2} \cos^{2}\phi
                + R_{y}^{\p\, 2}\sin^{2}\phi
                + \beta_{t}^{2}\Delta t^{2}, \\
 R_{\rm l}^{2} &=&R_{z}^{\p\, 2} \cos^2\theta
                + R_{x}^{\p\, 2} \sin^2\theta
                + \beta _{\rm l}^{2}\Delta t^{2}, \\
 R^{2}_{\rm ol} &=& (R_x^{\p\, 2}- R_z^{\p\, 2})
                     \cos\theta\sin\theta\cos\phi
                  +\beta _{t}\beta _{\rm l}\Delta t^{2}, \\
 R^{2}_{\rm os} &=& (R_{x}^{2}-R_{y}^{\p\, 2})
                     \cos \phi \sin \phi\,, \\
 R^{2}_{\rm sl} &=&(R_x^{\p\, 2} - R_z^{\p\, 2})
                    \cos\theta\sin\theta\sin\phi\,,
\end{eqnarray}
where an auxiliary quantity is introduced as
\begin{equation}
R_x^{2}=R_{x}^{\p\,2}\cos^2\theta + R_{z}^{\p\,2}\sin^2\theta\ .
\end{equation}
These results imply that all the radius parameters oscillate in 
the $K$ frame. In particular, a $\phi$ dependent oscillation 
appears in the radius parameters indexed either by the {\it side} 
or the {\it out} direction, as illustrated in Fig. 3. These 
oscillations are similar to those obtained in ref.~\cite{hhflow}, 
corresponding to $\theta =0$. We find that the radius parameters 
indexed by the {\it longitudinal} direction depend also on the 
zenithal angle $\theta$. A toy model for tilted ($\theta \ne 0$) 
ellipsoidal static pion sources was introduced in 
refs.~\cite{lisa}, to understand the $\phi$ dependent oscillations 
of measured HBT radii at AGS energies. In our case, the amplitude 
of the oscillations is reduced for heavier particles due to the 
hydrodynamic expansion, which results in a decrease of the lengths 
of homogeneity with increasing mass. 
The oscillations of the radius parameters were not related before 
either to hydrodynamic flow with ellipsoidal symmetry and tilt of 
the major axis or connected to the initial conditions of a 
hydrodynamic expansion. 

A check of the applicability of our hydrodynamic solution is that 
the BECF and the single particle spectrum become diagonal (after 
removing a term of $\beta_i\beta_j \Delta t^2$ from all the HBT
radius parameters) in the {\it same} frame, see 
eqs.~(\ref{e:ellsp},\ref{e:txp}-\ref{e:tzp}) and 
eqs.~(\ref{e:ellbecf},\ref{e:rxp}-\ref{e:rzp}). This frame is 
$K^\p$, the natural System of Ellipsoidal Expansion or SEE.

{\it Summary} ---  We have analytically evaluated the
single-particle spectrum and the two-particle Bose-Einstein
correlation function for a self-similarly expanding, exact,
ellipsoidal solution of the non-relativistic hydrodynamical
equations, assuming a constant freeze-out temperature. 

The parameters of the hydro solution at freeze-out time $T_f$,
$(X_f, Y_f, Z_f)$ and $(\dot X_f, \dot Y_f, \dot Z_f)$, can be
reconstructed from the measurement of the single particle 
spectrum and the two-particle correlation functions. The 
direction of the major axis of expansion in the center of mass 
frame of the collision is characterized by the polar angles 
$(\theta,0)$. With the exception of $R_{\rm l}$, all the radius 
parameters oscillate as function of $\phi$, while the radius 
parameters $R^2_{\rm l}\,$, $R^2_{\rm ol}\,$, $R_{\rm sl}^2$ 
depend also on $\theta$. All the radius parameters decrease with 
increasing mass, including all the cross terms. If $\theta\neq0$, 
the effective temperature in the transverse direction is 
increased by a contribution from the longitudinal expansion.

The initially more compressed longitudinal and impact parameter
directions (the $r_z$ and $r_x$ directions) expand more
dynamically~\cite{ellsol,nrtgen}, that implies 
$T_z^\p\geq T_x^\p\geq T_y^\p\,$. The initial time $t_0$ can be 
identified from the requirement that $\dot Y(t_0) = 0$. 
{\it The initial conditions for this hydrodynamical system can be 
uniquely reconstructed from final state measurements}. 
The function $\kappa (T)$ in the EoS influences only the time
evolution of the scales $(X,Y,Z)$ and temperature $T$. So, for a 
given EoS, one can uniquely reconstruct the initial conditions of 
hydrodynamic evolution from final ones.

We have deliberately chosen the presentation as simple as 
possible, which limits the direct applicability of our results 
in high-energy heavy-ion collisions only to sufficiently small 
transverse momentum, $p_{t}\ll m$, at mid-rapidity. But the 
scheme permits generalization in many points and still the 
qualitative features of our results may survive even in the 
relativistic regime. Generalizations to some relativistic flows
have been described in ref.~\cite{1drel}. Ref.~\cite{nrtgen}
includes another extension to an arbitrary, inhomogeneous,
ellipsoidally symmetric initial temperature profile, which does 
not change the time evolution of the scale parameters. It turns 
out that each of these generalizations is essentially 
straightforward. 

Although we considered a non-relativisitic problem, our results 
provide generic insight into the time evolution of non-central 
heavy-ion collisions and relate initial conditions to final-state 
observables in a simple and straightforward manner impossible 
before.

{\it Acknowledgments:} This work has been supported by the grants
FAPESP 00/04422-7, 99/09113-3, 01/05122-0 of S\~ao Paulo, Brazil,
by the Hungarian - Ukrainian S\&T grant 45014 (2M/125-199), the
OTKA grant T026435 and the NWO - OTKA  grant N 25487.
Y. S. and S. A. have been supported by  CNRS and the DLR grant 
2M-141- 2000.

\begin{figure}[tbp]
\vspace*{5.0cm}
\begin{center}
\includegraphics{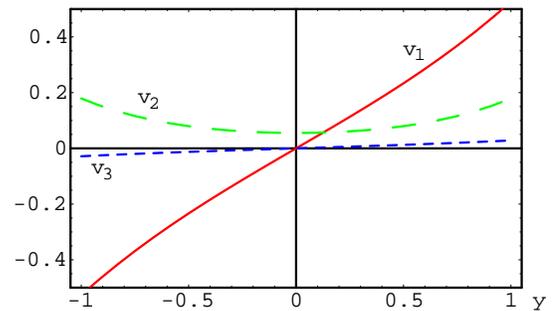}
\end{center}
\vspace{-1.3cm}
\caption{The directed, elliptic and third flows $v_1$, $v_2$, $v_3$
are illustrated, respectively with solid, long-dashed and
short-dashed lines, as a function of rapidity, for $m$ = 940 MeV,
$T_x^\p = 200$ MeV, $T_y^\p = 150$ MeV, $T_z^\p =700$ MeV, at a
fixed $k_t = 500$ MeV  and $\theta = \pi/5$, see
eqs.~(\ref{e:v1}-\ref{e:v3}).}
\end{figure}
\begin{figure}[tbp]
\vspace*{6.4cm}
\begin{center}
\includegraphics{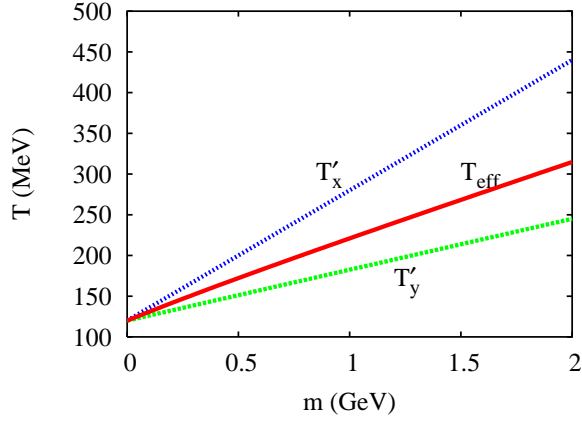}
\end{center}
\vspace{-1.5cm}
\caption{The linear mass dependence of the effective temperatures
in the transverse directions, $T_x^\p$, $T_y^\p$ and their
(harmonic) average $T_{\rm eff}$ for non-central heavy-ion 
collisions, if $T_f = 120$ MeV, $\dot X = 0.4$, and 
$\dot Y = 0.25$ and $\theta = 0$.}
\end{figure}
\begin{figure}[tbp]
\vspace*{6.cm}
\begin{center}
\includegraphics{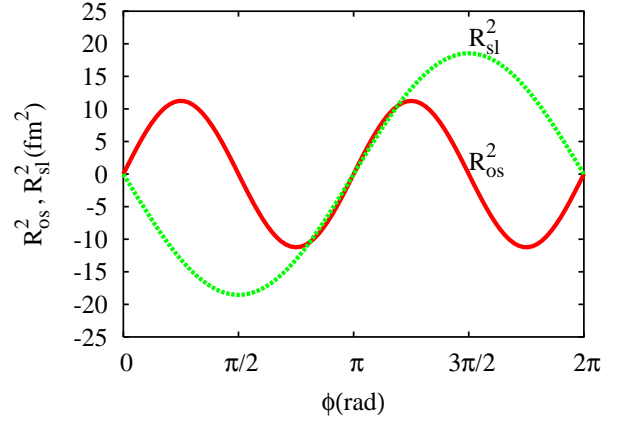}
\end{center}
\vspace{-1.5cm} \caption{ The out-side and the side-long cross
terms are plotted as a function of the polar angle $\phi$, for
$R_x^\p = 5$ fm, $R_y^\p = 4$ fm, $R_z^\p$ = 8 fm and $\theta  =
\pi/5$. Note that $R^2_{\rm sl} \propto \sin(\phi)$ while
$R^2_{\rm os} \propto \sin(2\phi)$. See eqs. (39) and (40).}
\end{figure}
\end{multicols}
\end{document}